\documentclass[fleqn]{llncs}
\usepackage{version}
\usepackage{funn}

\title{Functional Units for Natural Numbers}
\author{J.A. Bergstra \and C.A. Middelburg }
\institute{Informatics Institute, Faculty of Science,
           University of Amsterdam, \\
           Science Park~107, 1098~XG Amsterdam, the Netherlands \\
           \email{J.A.Bergstra@uva.nl,C.A.Middelburg@uva.nl}}

\begin{document}
\maketitle

\begin{abstract}
Interaction with services provided by an execution environment forms
part of the behaviours exhibited by instruction sequences under
execution.
Mechanisms related to the kind of interaction in question have been
proposed in the setting of thread algebra.
Like thread, service is an abstract behavioural concept.
The concept of a functional unit is similar to the concept of a service,
but more concrete.
A state space is inherent in the concept of a functional unit, whereas
it is not inherent in the concept of a service.
In this paper, we establish the existence of a universal computable
functional unit for natural numbers and related results.
\begin{classcode}
F.1.1, F.4.1.
\end{classcode}
\end{abstract}

\section{Introduction}
\label{sect-intro}

We take the view that sequential programs are in essence sequences of
instructions, and that interaction with services provided by an
execution environment forms part of the behaviours exhibited by
instruction sequences under execution (see e.g.~\cite{BL02a,PZ06a}).
The interaction in question is concerned with the processing of
instructions.
In earlier work, mechanisms that have a direct bearing on this kind of
interaction have been proposed in the setting of basic thread algebra
(see e.g.~\cite{BM09k,BP02a}).
Both thread and service are abstract behavioural concepts.

We experienced recently limitations of the concept of a service because
a state space is not inherent in this concept.
This forms the greater part of our motivation for introducing and
studying the concept of a functional unit in this paper.
This concept is similar to the concept of a service, but it is at a
lower level of abstraction.
In the concept of a functional unit, a state space is inherent.
Rather than first considering functional units in general for an
arbitrary state space, we first consider the special case where the
state space is the set of natural numbers.
This case is arguably the simplest significant case.
We establish general results concerning functional units for natural
numbers.
The main result is the existence of a universal computable functional
unit for natural numbers.
Results like this one are outside the scope of the concept of a service.

The work presented in this paper belongs to a line of research whose
working hypothesis is that instruction sequence is a central notion of
computer science.
In this line of research, program algebra~\cite{BL02a} is the setting
used for investigating issues in which instruction sequences are
involved.
Instruction sequences are also involved in the issues concerning
functional units investigated in this paper.
The starting-point of program algebra is the perception of a program as
a single-pass instruction sequence, i.e.\ a finite or infinite sequence
of instructions of which each instruction is executed at most once and
can be dropped after it has been executed or jumped over.
This perception is simple, appealing, and links up with practice.
Moreover, basic thread algebra~\cite{BL02a} is the setting used for
modelling the behaviours exhibited by instruction sequences under
execution.%
\footnote
{In~\cite{BL02a}, basic thread algebra is introduced under the name
 basic polarized process algebra.
}
In this paper, we use a program notation rooted in program algebra,
instead of program algebra itself.

This paper is organized as follows.
First, we give a survey of the program notation used in this paper
(Section~\ref{sect-PGLBbt}) and define its semantics using basic thread
algebra (Section~\ref{sect-BTAbt}).
Next, we extend basic thread algebra with operators that are related to
the processing of instructions by services (Section~\ref{sect-TSI}).
Then, we introduce the concept of a functional unit and related concepts
(Section~\ref{sect-func-unit}).
After that, we investigate functional units for natural numbers
(Section~\ref{sect-func-unit-nat}).
We also make some remarks about functional units for finite state
spaces (Section~\ref{sect-func-unit-fin}).
Finally, we make some concluding remarks (Section~\ref{sect-concl}).

\section{\PGLB\ with Boolean Termination}
\label{sect-PGLBbt}

In this section, we give a survey of the program notation \PGLBbt.
This program notation is a variant of the program notation \PGLB, which
belongs to a hierarchy of program notations rooted in program algebra
presented in~\cite{BL02a}.
\PGLBbt\ is \PGLB\ with the Boolean termination instructions $\haltP$
and $\haltN$ from~\cite{BM09k} instead of the termination instruction
$\halt$ from~\cite{BL02a}.
\PGLB\ and \PGLBbt\ are close to existing assembly languages and have
relative jump instructions.

In \PGLBbt, it is assumed that a fixed but arbitrary non-empty finite
set $\BInstr$ of \emph{basic instructions} has been given.
The intuition is that the execution of a basic instruction may modify a
state and produces $\True$ or $\False$ at its completion.

\PGLBbt\ has the following primitive instructions:
\begin{itemize}
\item
for each $a \in \BInstr$, a \emph{plain basic instruction} $a$;
\item
for each $a \in \BInstr$, a \emph{positive test instruction} $\ptst{a}$;
\item
for each $a \in \BInstr$, a \emph{negative test instruction} $\ntst{a}$;
\item
for each $l \in \Nat$, a \emph{forward jump instruction}
$\fjmp{l}$;
\item
for each $l \in \Nat$, a \emph{backward jump instruction}
$\bjmp{l}$;
\item
a \emph{positive termination instruction} $\haltP$;
\item
a \emph{negative termination instruction} $\haltN$.
\end{itemize}
\PGLBbt\ instruction sequences have the form
$u_1 \conc \ldots \conc u_k$, where $u_1,\ldots,u_k$ are primitive
instructions of \PGLBbt.

On execution of a \PGLBbt\ instruction sequence, these primitive
instructions have the following effects:
\begin{itemize}
\item
the effect of a positive test instruction $\ptst{a}$ is that basic
instruction $a$ is executed and execution proceeds with the next
primitive instruction if $\True$ is produced and otherwise the next
primitive instruction is skipped and execution proceeds with the
primitive instruction following the skipped one -- if there is no
primitive instruction to proceed with, deadlock occurs;
\item
the effect of a negative test instruction $\ntst{a}$ is the same as the
effect of $\ptst{a}$, but with the role of the value produced reversed;
\item
the effect of a plain basic instruction $a$ is the same as the effect of
$\ptst{a}$,\linebreak[2] but execution always proceeds as if $\True$ is
produced;
\item
the effect of a forward jump instruction $\fjmp{l}$ is that execution
proceeds with the $l$th next primitive instruction -- if $l$ equals $0$
or there is no primitive instruction to proceed with, deadlock occurs;
\item
the effect of a backward jump instruction $\bjmp{l}$ is that execution
proceeds with\linebreak[2] the $l$th previous primitive instruction --
if $l$ equals $0$ or there is no primitive instruction to
proceed with, deadlock occurs;
\item
the effect of the positive termination instruction $\haltP$ is that
execution terminates and in doing so delivers the Boolean value $\True$;
\item
the effect of the negative termination instruction $\haltP$ is that
execution terminates and in doing so delivers the Boolean value
$\False$.
\end{itemize}

\section{Thread Extraction}
\label{sect-BTAbt}

In this section, we make precise in the setting of \BTAbt\ (Basic Thread
Algebra with Boolean termination) which behaviours are exhibited on
execution by \PGLBbt\ instruction sequences.
We start by reviewing \BTAbt.

In \BTAbt, it is assumed that a fixed but arbitrary non-empty finite set
$\BAct$ of \emph{basic actions}, with $\Tau \not\in \BAct$, has been
given.
We write $\BActTau$ for $\BAct \union \set{\Tau}$.
The members of $\BActTau$ are referred to as \emph{actions}.

A thread is a behaviour which consists of performing actions in a
sequential fashion.
Upon each basic action performed, a reply from an execution environment
determines how it proceeds.
The possible replies are the Boolean values $\True$ (standing for true)
and $\False$ (standing for false).
Performing the action $\Tau$ leads always to the reply $\True$.

\BTAbt\ has one sort: the sort $\Thr$ of \emph{threads}.
We make this sort explicit because we will extend \BTAbt\ with
additional sorts in Section~\ref{sect-TSI}.
To build terms of sort $\Thr$, \BTAbt\ has the following constants and
operators:
\begin{itemize}
\item
the \emph{deadlock} constant $\const{\DeadEnd}{\Thr}$;
\item
the \emph{positive termination} constant $\const{\StopP}{\Thr}$;
\item
the \emph{negative termination} constant $\const{\StopN}{\Thr}$;
\item
for each $a \in \BActTau$, the binary \emph{postconditional composition}
operator $\funct{\pcc{\ph}{a}{\ph}}{\linebreak[2]\Thr \x \Thr}{\Thr}$.
\end{itemize}
We assume that there is a countably infinite set of variables of sort
$\Thr$ which includes $x,y,z$.
Terms of sort $\Thr$ are built as usual.
We use infix notation for postconditional composition.
We introduce \emph{action prefixing} as an abbreviation: $a \bapf p$,
where $p$ is a term of sort $\Thr$, abbreviates $\pcc{p}{a}{p}$.

The thread denoted by a closed term of the form $\pcc{p}{a}{q}$ will
first perform $a$, and then proceed as the thread denoted by $p$
if the reply from the execution environment is $\True$ and proceed as
the thread denoted by $q$ if the reply from the execution environment is
$\False$.
The threads denoted by $\DeadEnd$, $\StopP$ and $\StopN$ will become
inactive, terminate with Boolean value $\True$ and terminate with
Boolean value $\False$, respectively.

\BTAbt\ has only one axiom.
This axiom is given in Table~\ref{axioms-BTAbt}.%
\begin{table}[!tb]
\caption{Axiom of \BTAbt}
\label{axioms-BTAbt}
\begin{eqntbl}
\begin{axcol}
\pcc{x}{\Tau}{y} = \pcc{x}{\Tau}{x}                     & \axiom{T1}
\end{axcol}
\end{eqntbl}
\end{table}

Each closed \BTAbt\ term of sort $\Thr$ denotes a thread that will
become inactive or terminate after it has performed finitely many
actions.
Infinite threads can be described by linear recursion.
A \emph{linear recursive specification} over \BTAbt\ is a set of
recursion equations $E = \set{x = t_x \where x \in V}$, where $V$ is a
set of variables of sort $\Thr$ and each $t_x$ is a \BTAbt\ term of the
form $\DeadEnd$, $\StopP$, $\StopN$ or $\pcc{y}{a}{z}$ with $y,z \in V$.
We are only interested in models of \BTAbt\ in which linear recursive
specifications have unique solutions.
Regular threads, i.e.\ threads that can only be in a finite number of
states, are solutions of finite linear recursive specifications.

To reason about infinite threads, we assume the infinitary conditional
equation AIP (Approximation Induction Principle).
AIP is based on the view that two threads are identical if their
approximations up to any finite depth are identical.
The approximation up to depth $n$ of a thread is obtained by cutting it
off after it has performed $n$ actions.
In AIP, the approximation up to depth $n$ is phrased in terms of the
unary \emph{projection} operator $\funct{\projop{n}}{\Thr}{\Thr}$.
AIP and the axioms for the projection operators are given in
Table~\ref{axioms-AIP}.%
\begin{table}[!tb]
\caption{Approximation induction principle}
\label{axioms-AIP}
\begin{eqntbl}
\begin{axcol}
\AND{n \geq 0}{} \proj{n}{x} = \proj{n}{y} \Implies
                                                  x = y & \axiom{AIP} \\
\proj{0}{x} = \DeadEnd                                  & \axiom{P0} \\
\proj{n+1}{\StopP} = \StopP                             & \axiom{P1a} \\
\proj{n+1}{\StopN} = \StopN                             & \axiom{P1b} \\
\proj{n+1}{\DeadEnd} = \DeadEnd                         & \axiom{P2} \\
\proj{n+1}{\pcc{x}{a}{y}} =
                      \pcc{\proj{n}{x}}{a}{\proj{n}{y}} & \axiom{P3}
\end{axcol}
\end{eqntbl}
\end{table}
In this table, $a$ stands for an arbitrary action from $\BActTau$ and
$n$ stands for an arbitrary natural number.

The behaviours exhibited on execution by \PGLBbt\ instruction sequences
are considered to be regular threads, with the basic instructions taken
for basic actions.
The \emph{thread extraction} operation $\extr{\ph}$ defines, for each
\PGLBbt\ instruction sequence, the behaviour exhibited on execution by
that \PGLBbt\ instruction sequence.
The thread extraction operation is defined by
$\extr{u_1 \conc \ldots \conc u_k} =
 \extr{1,u_1 \conc \ldots \conc u_k}$,
where the auxiliary operation $\extr{\ph,\ph}$ is defined by the
equations given in Table~\ref{axioms-thread-extr} (for $a \in \BInstr$
and $l,i \in \Nat$)%
\begin{table}[!t]
\caption{Defining equations for thread extraction operation}
\label{axioms-thread-extr}
\begin{eqntbl}
\begin{aceqns}
\extr{i,u_1 \conc \ldots \conc u_k} & = & \DeadEnd
& \mif \mathrm{not}\; 1 \leq i \leq k \\
\extr{i,u_1 \conc \ldots \conc u_k} & = &
a \bapf \extr{i+1,u_1 \conc \ldots \conc u_k}
& \mif u_i = a \\
\extr{i,u_1 \conc \ldots \conc u_k} & = &
\pcc{\extr{i+1,u_1 \conc \ldots \conc u_k}}{a}
    {\extr{i+2,u_1 \conc \ldots \conc u_k}}
& \mif u_i = +a \\
\extr{i,u_1 \conc \ldots \conc u_k} & = &
\pcc{\extr{i+2,u_1 \conc \ldots \conc u_k}}{a}
    {\extr{i+1,u_1 \conc \ldots \conc u_k}}
& \mif u_i = -a \\
\extr{i,u_1 \conc \ldots \conc u_k} & = &
\extr{i+l,u_1 \conc \ldots \conc u_k}
& \mif u_i = \fjmp{l} \\
\extr{i,u_1 \conc \ldots \conc u_k} & = &
\extr{i \monus l,u_1 \conc \ldots \conc u_k}
& \mif u_i = \bjmp{l} \\
\extr{i,u_1 \conc \ldots \conc u_k} & = & \StopP
& \mif u_i = \haltP \\
\extr{i,u_1 \conc \ldots \conc u_k} & = & \StopN
& \mif u_i = \haltN
\end{aceqns}
\end{eqntbl}
\end{table}
and the rule that $\extr{i,u_1 \conc \ldots \conc u_k} = \DeadEnd$ if
$u_i$ is the beginning of an infinite jump chain.%
\footnote
{This rule can be formalized, cf.~\cite{BM07g}.}

\section{Interaction between Threads and Services}
\label{sect-TSI}

A thread may perform a basic action for the purpose of requesting a
named service to process a method and to return a reply value at
completion of the processing of the method.
In this section, we extend \BTAbt\ such that this kind of interaction
between threads and services can be dealt with, resulting in \TAbt.
This involves the introduction of service families: collections of named
services.

It is assumed that a fixed but arbitrary non-empty finite set $\Meth$ of
\emph{methods} has been given.
Methods play the role of commands.
A service is able to process certain methods.
The processing of a method by a service may involve a change of state of
the service and at completion of the processing of the method the
service produces a reply value.
The set $\Replies$ of \emph{reply values} is the set
$\set{\True,\False,\Div}$.

In \SFA, the algebraic theory of service families introduced below, the
following is assumed with respect to services:
\begin{itemize}
\item
a set $\Services$ of services has been given together with:
\begin{itemize}
\item
for each $m \in \Meth$,
a total function $\funct{\effect{m}}{\Services}{\Services}$;
\item
for each $m \in \Meth$,
a total function $\funct{\sreply{m}}{\Services}{\Replies}$;
\end{itemize}
satisfying the condition that there exists a unique $S \in \Services$
with $\effect{m}(S) = S$ and $\sreply{m}(S) = \Div$ for all
$m \in \Meth$;
\item
a signature $\Sig{\Services}$ has been given that includes the following
sort:
\begin{itemize}
\item
the sort $\Serv$ of \emph{services};
\end{itemize}
and the following constant and operators:
\begin{itemize}
\item
the \emph{empty service} constant $\const{\emptyserv}{\Serv}$;
\item
for each $m \in \Meth$,
the \emph{derived service} operator
$\funct{\derive{m}}{\Serv}{\Serv}$;
\end{itemize}
\item
$\Services$ and $\Sig{\Services}$ are such that:
\begin{itemize}
\item
each service in $\Services$ can be denoted by a closed term of sort
$\Serv$;
\item
the constant $\emptyserv$ denotes the unique $S \in \Services$ such
that $\effect{m}(S) = S$ and $\sreply{m}(S) = \Div$ for all
$m \in \Meth$;
\item
if closed term $t$ denotes service $S$, then $\derive{m}(t)$ denotes
service $\effect{m}(S)$.
\end{itemize}
\end{itemize}

When a request is made to service $S$ to process method $m$:
\begin{itemize}
\item
if $\sreply{m}(S) \neq \Div$, then $S$ processes $m$, produces the reply
$\sreply{m}(S)$, and next proceeds as $\effect{m}(S)$;
\item
if $\sreply{m}(S) = \Div$, then $S$ rejects the request to process
method $m$.
\end{itemize}
The unique service $S$ such that $\effect{m}(S) = S$ and
$\sreply{m}(S) = \Div$ for all $m \in \Meth$ is called the \emph{empty}
service.
It is the service that is unable to process any method.

It is also assumed that a fixed but arbitrary non-empty finite set
$\Foci$ of \emph{foci} has been given.
Foci play the role of names of services in the service family offered by
an execution environment.
A service family is a set of named services where each name occurs only
once.

\SFA\ has the sorts, constants and operators in $\Sig{\Services}$
and in addition the following sort:
\begin{itemize}
\item
the sort $\ServFam$ of \emph{service families};
\end{itemize}
and the following constant and operators:
\begin{itemize}
\item
the \emph{empty service family} constant $\const{\emptysf}{\ServFam}$;
\item
for each $f \in \Foci$, the unary \emph{singleton service family}
operator $\funct{\mathop{f{.}} \ph}{\Serv}{\ServFam}$;
\item
the binary \emph{service family composition} operator
$\funct{\ph \sfcomp \ph}{\ServFam \x \ServFam}{\ServFam}$;
\item
for each $F \subseteq \Foci$, the unary \emph{encapsulation} operator
$\funct{\encap{F}}{\ServFam}{\ServFam}$.
\end{itemize}
We assume that there is a countably infinite set of variables of sort
$\ServFam$ which includes $u,v,w$.
Terms are built as usual in the many-sorted case
(see e.g.~\cite{ST99a,Wir90a}).
We use prefix notation for the singleton service family operators and
infix nota\-tion for the service family composition operator.

The service family denoted by $\emptysf$ is the empty service family.
The service family denoted by a closed term of the form $f.H$ consists
of one named service only, the service concerned is the service denoted
by $H$, and the name of this service is $f$.
The service family denoted by a closed term of the form $C \sfcomp D$
consists of all named services that belong to either the service family
denoted by $C$ or the service family denoted by $D$.
In the case where a named service from the service family denoted by $C$
and a named service from the service family denoted by $D$ have the same
name, they collapse to an empty service with the name concerned.
The service family denoted by a closed term of the form $\encap{F}(C)$
consists of all named services with a name not in $F$ that belong to the
service family denoted by $C$.

The service family composition operator takes the place of the
non-interfering combination operator from~\cite{BP02a}.
As suggested by the name, service family composition is composition of
service families.
Non-interfering combination is composition of services, which has the
disadvantage that its usefulness is rather limited without an additional
renaming mechanism.

The axioms of \SFA\ are given in Table~\ref{axioms-SFA}.%
\begin{table}[!t]
\caption{Axioms of \SFA}
\label{axioms-SFA}
\begin{eqntbl}
\begin{axcol}
u \sfcomp \emptysf = u                                 & \axiom{SFC1} \\
u \sfcomp v = v \sfcomp u                              & \axiom{SFC2} \\
(u \sfcomp v) \sfcomp w = u \sfcomp (v \sfcomp w)      & \axiom{SFC3} \\
f.H \sfcomp f.H' = f.\emptyserv                        & \axiom{SFC4}
\end{axcol}
\qquad
\begin{saxcol}
\encap{F}(\emptysf) = \emptysf                       & & \axiom{SFE1} \\
\encap{F}(f.H) = \emptysf & \mif f \in F               & \axiom{SFE2} \\
\encap{F}(f.H) = f.H      & \mif f \notin F            & \axiom{SFE3} \\
\encap{F}(u \sfcomp v) =
\encap{F}(u) \sfcomp \encap{F}(v)                    & & \axiom{SFE4}
\end{saxcol}
\end{eqntbl}
\end{table}
In this table, $f$ stands for an arbitrary focus from $\Foci$ and $H$
and $H'$ stand for arbitrary closed terms of sort $\Serv$.
The axioms of \SFA\ simply formalize the informal explanation given above.

Below we will introduce two operators related to the interaction between
threads and services.
They are called the apply operator and the reply operator.
The apply operator is concerned with the effects of threads on service
families and therefore produces service families.
The reply operator is concerned with the effects of service families on
the Boolean values that threads deliver at their termination.
The reply operator does not only produce Boolean values: it produces a
special value in cases where no termination takes place.

For the set $\BAct$ of basic actions, we take the set
$\set{f.m \where f \in \Foci, m \in \Meth}$.
Both operators mentioned above relate to the processing of methods by
services from a service family in pursuance of basic actions performed
by a thread.
The service involved in the processing of a method is the service whose
name is the focus of the basic action in question.

\TAbt\ has the sorts, constants and operators of both \BTAbt\ and \SFA,
and in addition the following sort:
\begin{itemize}
\item
the sort $\Repl$ of \emph{replies};
\end{itemize}
and the following constants and operators:
\begin{itemize}
\item
the \emph{reply} constants $\const{\True,\False,\Div}{\Repl}$;
\item
the binary \emph{apply} operator
$\funct{\ph \sfapply \ph}{\Thr \x \ServFam}{\ServFam}$;
\item
the binary \emph{reply} operator
$\funct{\ph \sfreply \ph}{\Thr \x \ServFam}{\Repl}$.
\end{itemize}
We use infix notation for the apply and reply operators.

The service family denoted by a closed term of the form $p \sfapply C$
and the reply denoted by a closed term of the form $p \sfreply C$ are
the service family and reply, respectively, that result from processing
the method of each basic action with a focus of the service family
denoted by $C$ that the thread denoted by $p$ performs, where the
processing is done by the service in that service family with the focus
of the basic action as its name.
When the method of a basic action performed by a thread is processed by
a service, the service changes in accordance with the method concerned,
and affects the thread as follows: the two ways to proceed reduces to
one on the basis of the reply value produced by the service.
The reply is the Boolean value that the thread denoted by $p$ delivers
at termination if it terminates and the value $\Div$ (standing for
divergent) if it does not terminate.

The axioms of \TAbt\ are the axioms of \BTAbt, the axioms of \SFA, and
the axioms given in Tables~\ref{axioms-apply} and~\ref{axioms-reply}.%
\begin{table}[!t]
\caption{Axioms for apply operator}
\label{axioms-apply}
\begin{eqntbl}
\begin{saxcol}
\StopP \sfapply u = u                                  & & \axiom{A1} \\
\StopN \sfapply u = u                                  & & \axiom{A2} \\
\DeadEnd \sfapply u = \emptysf                         & & \axiom{A3} \\
(\Tau \bapf x) \sfapply u = x \sfapply u               & & \axiom{A4} \\
(\pcc{x}{f.m}{y}) \sfapply \encap{\set{f}}(u) = \emptysf
                                                       & & \axiom{A5} \\
(\pcc{x}{f.m}{y}) \sfapply (f.H \sfcomp \encap{\set{f}}(u)) =
x \sfapply (f.\derive{m}H \sfcomp \encap{\set{f}}(u))
                           & \mif \sreply{m}(H) = \True  & \axiom{A6} \\
(\pcc{x}{f.m}{y}) \sfapply (f.H \sfcomp \encap{\set{f}}(u)) =
y \sfapply (f.\derive{m}H \sfcomp \encap{\set{f}}(u))
                           & \mif \sreply{m}(H) = \False & \axiom{A7} \\
(\pcc{x}{f.m}{y}) \sfapply (f.H \sfcomp \encap{\set{f}}(u)) = \emptysf
                           & \mif \sreply{m}(H) = \Div   & \axiom{A8} \\
\AND{n \geq 0}{} \proj{n}{x} \sfapply u = \proj{n}{y} \sfapply v
                 \Implies x \sfapply u = y  \sfapply v & & \axiom{A9}
\end{saxcol}
\end{eqntbl}
\end{table}%
\begin{table}[!t]
\caption{Axioms for reply operator}
\label{axioms-reply}
\begin{eqntbl}
\begin{saxcol}
\StopP \sfreply u = \True                              & & \axiom{R1} \\
\StopN \sfreply u = \False                             & & \axiom{R2} \\
\DeadEnd \sfreply u = \Div                             & & \axiom{R3} \\
(\Tau \bapf x) \sfreply u = x \sfreply u               & & \axiom{R4} \\
(\pcc{x}{f.m}{y}) \sfreply \encap{\set{f}}(u) = \Div   & & \axiom{R5} \\
(\pcc{x}{f.m}{y}) \sfreply (f.H \sfcomp \encap{\set{f}}(u)) =
x \sfreply (f.\derive{m}H \sfcomp \encap{\set{f}}(u))
                           & \mif \sreply{m}(H) = \True  & \axiom{R6} \\
(\pcc{x}{f.m}{y}) \sfreply (f.H \sfcomp \encap{\set{f}}(u)) =
y \sfreply (f.\derive{m}H \sfcomp \encap{\set{f}}(u))
                           & \mif \sreply{m}(H) = \False & \axiom{R7} \\
(\pcc{x}{f.m}{y}) \sfreply (f.H \sfcomp \encap{\set{f}}(u)) = \Div
                           & \mif \sreply{m}(H) = \Div   & \axiom{R8} \\
\AND{n \geq 0}{} \proj{n}{x} \sfreply u = \proj{n}{y} \sfreply v
                 \Implies x \sfreply u = y  \sfreply v & & \axiom{R9}
\end{saxcol}
\end{eqntbl}
\end{table}
In these tables, $f$ stands for an arbitrary focus from $\Foci$, $m$
stands for an arbitrary method from $\Meth$, $H$ stands for an arbitrary
term of sort $\Serv$, and $n$ stands for an arbitrary natural number.
The axioms simply formalize the informal explanation given above and in
addition stipulate what is the result of apply and reply if
inappropriate foci or methods are involved.
Axioms A9 and R9 allow for reasoning about infinite threads in the
contexts of apply and reply, respectively.

\section{Functional Units}
\label{sect-func-unit}

In this section, we introduce the concept of a functional unit and
related concepts such as a functional unit degree.

It is assumed that a non-empty set $\FUS$ of \emph{states} has been
given.
As before, it is assumed that a non-empty finite set $\MN$ of methods
has been given.
However, in the setting of functional units, methods serve as names of
operations on a state space.
For that reason, the members of $\MN$ will henceforth be called
\emph{method names}.

A \emph{method operation} on $\FUS$ is a total function from $\FUS$ to
$\Bool \x \FUS$.
A \emph{partial method operation} on $\FUS$ is a partial function from
$\FUS$ to $\Bool \x \FUS$.
We write $\MO(\FUS)$ for the set of all method operations on $\FUS$.
We write $M^r$ and $M^e$, where $M \in \MO(\FUS)$, for the unique
functions $\funct{R}{\FUS}{\Bool}$ and $\funct{E}{\FUS}{\FUS}$,
respectively, such that $M(s) = \tup{R(s),E(s)}$ for all $s \in \FUS$.

A \emph{functional unit} for $\FUS$ is a finite subset $\cH$ of
$\MN \x \MO(\FUS)$ such that \mbox{$\tup{m,M} \in \cH$} and
$\tup{m,M'} \in \cH$ implies $M = M'$.
We write $\FU(\FUS)$ for the set of all functional units for $\FUS$.
We write $\IF(\cH)$, where $\cH \in \FU(\FUS)$, for the set
$\set{m \in \MN \where \Exists{M \in \MO(\FUS)}{\tup{m,M} \in \cH}}$.
We write $m_\cH$, where $\cH \in \FU(\FUS)$ and $m \in \IF(\cH)$, for
the unique $M \in \MO(\FUS)$ such that $\tup{m,M} \in \cH$.

We look upon the set $\IF(\cH)$, where $\cH \in \FU(\FUS)$, as the
interface of $\cH$.
It looks to be convenient to have a notation for the restriction of a
functional unit to a subset of its interface.
We write $\tup{I,\cH}$, where $\cH \in \FU(\FUS)$ and
$I \subseteq \IF(\cH)$, for the functional unit
$\set{\tup{m,M} \in \cH \where m \in I}$.

Let $\cH \in \FU(\FUS)$.
Then an \emph{extension} of $\cH$ is an $\cH' \in \FU(\FUS)$ such that
$\cH \subseteq \cH'$.

The following is a simple illustration of the use of functional units.
An unbounded counter can be modelled by a functional unit for $\Nat$
with method operations for set to zero, increment by one, decrement by
one, and test on zero.

According to the definition of a functional unit,
$\emptyset \in \FU(\FUS)$.
By that we have a unique functional unit with an empty interface, which
is not very interesting in itself.
However, when considering services that behave according to functional
units, $\emptyset$ is exactly the functional unit according to which the
empty service $\emptyserv$ (the service that is not able to process any
method) behaves.

The method names attached to method operations in functional units
should not be confused with the names used to denote specific method
operations in describing functional units.
Therefore, we will comply with the convention to use names beginning
with a lower-case letter in the former case and names beginning with an
upper-case letter in the latter case.

We will use \PGLBbt\ instruction sequences to derive partial method
operations from the method operations of a functional unit.
We write $\Lf{I}$, where $I \subseteq \MN$, for the set of all \PGLBbt\
instruction sequences, taking the set $\set{f.m \where m \in I}$ as the
set $\BInstr$ of basic instructions.

The derivation of partial method operations from the method operations
of a functional unit involves services whose processing of methods
amounts to replies and service changes according to corresponding method
operations of the functional unit concerned.
These services can be viewed as the behaviours of a machine, on which
the processing in question takes place, in its different states.
We take the set $\FU(\FUS) \x \FUS$ as the set $\Services$ of services.
We write $\cH(s)$, where $\cH \in \FU(\FUS)$ and $s \in \FUS$, for the
service $\tup{\cH,s}$.
The functions $\effect{m}$ and $\sreply{m}$ are defined as follows:
\begin{ldispl}
\begin{aeqns}
\effect{m}(\cH(s)) & = &
\Biggl\{
\begin{array}[c]{@{}l@{\;\;}l@{}}
\cH(m_\cH^e(s))            & \mif m \in \IF(\cH) \\
{\emptyset}(s')            & \mif m \notin \IF(\cH)\;,
\end{array}
\beqnsep
\sreply{m}(\cH(s))  & = &
\Biggl\{
\begin{array}[c]{@{}l@{\;\;}l@{}}
m_\cH^r(s) \phantom{\cH()} & \mif m \in \IF(\cH) \\
\Div                       & \mif m \notin \IF(\cH)\;,
\end{array}
\end{aeqns}
\end{ldispl}
where $s'$ is a fixed but arbitrary state in $S$.
We assume that each $\cH(s) \in \Services$ can be denoted by a closed
term of sort $\Serv$.
In this connection, we use the following notational convention: for each
$\cH(s) \in \Services$, we write $\cterm{\cH(s)}$ for an arbitrary
closed term of sort $\Thr$ that denotes $\cH(s)$.
The ambiguity thus introduced could be obviated by decorating $\cH(s)$
wherever it stands for a closed term.
However, in this paper, it is always immediately clear from the context
whether it stands for a closed term.
Moreover, we believe that the decorations are more often than not
distracting.
Therefore, we leave it to the reader to make the decorations mentally
wherever appropriate.

Let $\cH \in \FU(\FUS)$, and let $I \subseteq \IF(\cH)$.
Then an instruction sequence $x \in \Lf{I}$ produces a partial method
operation $\moextr{x}{\cH}$ as follows:
\begin{ldispl}
\begin{aceqns}
\moextr{x}{\cH}(s) & = &
\tup{\moextr{x}{\cH}^r(s),\moextr{x}{\cH}^e(s)}
 & \mif \moextr{x}{\cH}^r(s) = \True \Or
        \moextr{x}{\cH}^r(s) = \False\;, \\
\moextr{x}{\cH}(s) & \mathrm{is} & \mathrm{undefined}
 & \mif \moextr{x}{\cH}^r(s) = \Div\;,
\end{aceqns}
\end{ldispl}
where
\begin{ldispl}
\begin{aeqns}
\moextr{x}{\cH}^r(s) & = & x \sfreply f.\cterm{\cH(s)}\;, \\
\moextr{x}{\cH}^e(s) & = &
\mathrm{the\;unique}\; s' \in S\; \mathrm{such\;that}\;
 x \sfapply f.\cterm{\cH(s)} = f.\cterm{\cH(s')}\;.
\end{aeqns}
\end{ldispl}
If $\moextr{x}{\cH}$ is total, then it is called a
\emph{derived method operation} of $\cH$.

The binary relation $\below$ on $\FU(\FUS)$ is defined by
$\cH \below \cH'$ iff for all $\tup{m,M} \in \cH$, $M$ is a derived
method operation of $\cH'$.
The binary relation $\equiv$ on $\FU(\FUS)$ is defined by
$\cH \equiv \cH'$ iff $\cH \below \cH'$ and $\cH' \below \cH$.

\begin{theorem}
\label{theorem-below-equiv}
\mbox{}
\begin{enumerate}
\item
$\below$ is transitive;
\item
$\equiv$ is an equivalence relation.
\end{enumerate}
\end{theorem}
\begin{proof}
Property~1:
We have to prove that $\cH \below \cH'$ and $\cH' \below \cH''$ implies
$\cH \below \cH''$.
It is sufficient to show that we can obtain instruction sequences in
$\Lf{\IF(\cH'')}$ that produce the method operations of $\cH$ from the
instruction sequences in $\Lf{\IF(\cH')}$ that produce the method
operations of $\cH$ and the instruction sequences in $\Lf{\IF(\cH'')}$
that produce the method operations of $\cH'$.
Without loss of generality, we may assume that all instruction sequences
are of the form $u_1 \conc \ldots \conc u_k \conc \haltP \conc \haltN$,
where, for each $i \in [1,k]$, $u_i$ is a positive test instruction, a
forward jump instruction or a backward jump instruction.
Let $m \in \IF(\cH)$,
let $M$ be such that $\tup{m,M} \in \cH$, and
let $x_m \in \Lf{\IF(\cH')}$ be such that $M = \moextr{x_m}{\cH'}$.
Suppose that $\IF(\cH') = \set{m'_1,\ldots,m'_n}$.
For each $i \in [1,n]$,
let $M'_i$ be such that $\tup{m'_i,M'_i} \in \cH'$ and
let
$x_{m'_i} = u_1^i \conc \ldots \conc u_{k_i}^i \conc \haltP \conc \haltN
  \in \Lf{\IF(\cH'')}$ be such that $M'_i = \moextr{x_{m'_i}}{\cH''}$.
Consider the $x'_m \in \Lf{\IF(\cH'')}$ obtained from $x_m$ as follows:
for each $i \in [1,n]$,
(i)~first increase each jump over the leftmost occurrence of
$\ptst{f.m'_i}$ in $x_m$ with $k_i + 1$, and next replace this
instruction by $u_1^i \conc \ldots \conc u_{k_i}^i$;
(ii)~repeat the previous step as long as their are occurrences of
$\ptst{f.m'_i}$.
It is easy to see that $M = \moextr{x'_m}{\cH''}$.

Property~2:
It follows immediately from the definition of $\equiv$ that $\equiv$ is
symmetric and from the definition of $\below$ that $\below$ is
reflexive.
From these properties, Property~1 and the definition of $\equiv$, it
follows immediately that $\equiv$ is symmetric, reflexive and
transitive.
\qed
\end{proof}

The members of the quotient set $\FU(\FUS) \mdiv {\equiv}$ are called
\emph{functional unit degrees}.
Let $\cH \in \FU(\FUS)$ and $\cD \in \FU(\FUS) \mdiv {\equiv}$.
Then $\cD$ is a \emph{functional unit degree below}~$\cH$ if there
exists an $\cH' \in \cD$ such that $\cH' \below \cH$.

\section{Functional Units for Natural Numbers}
\label{sect-func-unit-nat}

In this section, we investigate functional units for natural numbers.
The main consequences of considering the special case where the state
space is $\Nat$ are the following:
(i)~$\Nat$ is infinite,
(ii)~there is a notion of computability known which can be used without
further preparations.

An example of a functional unit in $\FU(\Nat)$ is an unbounded counter.
The method names involved are $\setzero$, $\incr$, $\decr$, and
$\iszero$.
The method operations involved are the functions $\Setzero$, $\Incr$,
$\Decr$, $\funct{\Iszero}{\Nat}{\Bool \x \Nat}$ defined as follows:
\begin{ldispl}
\begin{aeqns}
\Setzero(x) & = & \tup{\True,0}\;,     \\
\Incr(x)    & = & \tup{\True,x + 1}\;, \\
\Decr(x)    & = &
\Biggl\{
\begin{array}[c]{@{}l@{\;\;}l@{}}
\tup{\True,x - 1} & \mif x > 0\;, \\
\tup{\False,0}    & \mif x = 0\;,
\end{array}
\beqnsep
\Iszero(x) & = &
\Biggl\{
\begin{array}[c]{@{}l@{\;\;}l@{}}
\tup{\True,x} \!\phantom{{} - 1} & \mif x = 0\;, \\
\tup{\False,x}                   & \mif x > 0\;.
\end{array}
\end{aeqns}
\end{ldispl}
The functional unit $\Counter$ is defined as follows:
\begin{ldispl}
\Counter =
\set{\tup{\setzero,\Setzero},\tup{\incr,\Incr},
     \tup{\decr,\Decr},\tup{\iszero,\Iszero}}\;.
\end{ldispl}

\begin{proposition}
\label{prop-inf-many-degrees}
\sloppy
There are infinitely many functional unit degrees below
$\tup{\set{\decr,\iszero},\Counter}$.
\end{proposition}
\begin{proof}
For each $n \in \Nat$, we define a functional unit $\cH_n \in \FU(\Nat)$
such that $\cH_n \leq \tup{\set{\decr,\iszero},\Counter}$ as follows:
\begin{ldispl}
\cH_n = \set{\tup{\decrn{n},\Decrn{n}},\tup{\iszero,\Iszero}}\;,
\end{ldispl}
where
\begin{ldispl}
\Decrn{n}(x) =
\biggl\{
\begin{array}[c]{@{}l@{\;\;}l@{}}
\tup{\True, x - n} & \mif x \geq n \\
\tup{\False,0}     & \mif x < n\;.
\end{array}
\end{ldispl}
Let $n,m \in \Nat$ be such that $n < m$.
Then $\Decrn{n}(m) = \tup{\True,m - n}$.
However, there does not exist an $x \in \Lf{\IF(\cH_m)}$ such that
$\moextr{x}{\cH_m}(m) = \tup{\True,m - n}$ because
$\Decrn{m}(m) = \tup{\True,0}$.
Hence, $\cH_n \not\leq \cH_m$ for all $n,m \in \Nat$ with $n < m$.
\qed
\end{proof}

A method operation $M \in \MO(\Nat)$ is \emph{computable} if there exist
computable functions $\funct{F,G}{\Nat}{\Nat}$ such that
$M(n) = \tup{\beta(F(n)),G(n)}$ for all $n \in \Nat$,
where $\funct{\beta}{\Nat}{\Bool}$ is inductively defined by
$\beta(0) = \True$ and $\beta(n + 1) = \False$.
A functional unit $\cH \in \FU(\Nat)$ is \emph{computable} if, for each
$\tup{m,M} \in \cH$, $M$ is computable.

\begin{theorem}
\label{theorem-computable}
Let $\cH,\cH' \in \FU(\Nat)$ be such that $\cH \below \cH'$.
Then $\cH$ is computable if $\cH'$ is computable.
\end{theorem}
\begin{proof}
We will show that all derived method operations of $\cH'$ are
computable.

Take an arbitrary $P \in \Lf{\IF(\cH')}$ such that $\moextr{P}{\cH'}$ is
a derived method operations of $\cH'$.
It follows immediately from the definition of thread extraction that
$\extr{P}$ is the solution of a finite linear recursive specification
over \BTAbt, i.e.\ a finite guarded recursive specification over \BTAbt\
in which the right-hand side of each equation is a \BTAbt\ term of the
form $\DeadEnd$, $\StopP$, $\StopN$ or $\pcc{x}{a}{y}$ where $x$ and $y$
are variables of sort $\Thr$.
Let $E$ be a finite linear recursive specification over \BTAbt\ of which
the solution for $x_1$ is $\extr{P}$.
Because $\moextr{P}{\cH'}$ is total, it may be assumed without loss of
generality that $\DeadEnd$ does not occur as the right-hand side of an
equation in $E$.
Suppose that
\begin{ldispl}
E =
\set{x_i = \pcc{x_{l(i)}}{f.m_i}{x_{r(i)}} \where i \in [1,n]} \union
\set{x_{n+1} = \StopP, x_{n+2} = \StopN}\;.
\end{ldispl}
From this set of equations, using the relevant axioms and definitions,
we obtain a set of equations of which the solution for $F_1$ is
$\moextr{P}{\cH'}^e$:
\begin{ldispl}
\set{F_i(s) = F_{l(i)}({m_i}_{\cH'}^e(s)) \mmul \nsg(\chi_i(s)) +
              F_{r(i)}({m_i}_{\cH'}^e(s)) \mmul  \sg(\chi_i(s))
      \where i \in [1,n]}
\\ \quad {} \union
\set{F_{n+1}(s) = s, F_{n+2}(s) = s}\;,
\end{ldispl}
where, for every $i \in [1,n]$, the function
$\funct{\chi_i}{\Nat}{\Nat}$ is such that for all $s \in \Nat$:
\begin{ldispl}
\chi_i(s) = 0 \;\Iff\; {m_i}_{\cH'}^r(s) = \True\;,
\end{ldispl}
and the functions $\funct{\sg,\nsg}{\Nat}{\Nat}$ are defined as usual:
\begin{ldispl}
\begin{aeqns}
\sg(0)     & = & 0\;, \\
\sg(n + 1) & = & 1\;,
\end{aeqns}
\qquad\qquad
\begin{aeqns}
\nsg(0)     & = & 1\;, \\
\nsg(n + 1) & = & 0\;.
\end{aeqns}
\end{ldispl}
It follows from the way in which this set of equations is obtained from
$E$, the fact that ${m_i}_{\cH'}^e$ and $\chi_i$ are computable for each
$i \in [1,n]$, and the fact that $\sg$ and $\nsg$ are computable, that
this set of equations is equivalent to a set of equations by which
$\moextr{P}{\cH'}^e$ is defined recursively in the sense of Kleene
(see~\cite{Kle36a}).
This means that $\moextr{P}{\cH'}^e$ is general recursive, and hence
computable.

In a similar way, it is proved that $\moextr{P}{\cH'}^r$ is computable.
\qed
\end{proof}

A computable $\cH \in \FU(\Nat)$ is \emph{universal} if for each
computable $\cL \in \FU(\Nat)$, we have $\cL \below \cH$.
There exists a universal computable functional unit for natural numbers.
\begin{theorem}
\label{theorem-universal-fu}
There exists a computable $\cH \in \FU(\Nat)$ that is universal.
\end{theorem}
\begin{proof}
We will show that there exists a computable $\cH \in \FU(\Nat)$ with the
property that each computable $M \in \MO(\Nat)$ is a derived method
operation of $\cH$.

As a corollary of Theorem~10.3 from~\cite{SS63a},%
\footnote{That theorem can be looked upon as a corollary of Theorem~Ia
          from~\cite{Min61a}.}
we have that each computable $M \in \MO(\Nat)$ can be computed by means
of a register machine with six registers, say $\rmreg{0}$, $\rmreg{1}$,
$\rmreg{2}$, $\rmreg{3}$, $\rmreg{4}$, and $\rmreg{5}$.
The registers are used as follows:
$\rmreg{0}$ as input register;
$\rmreg{1}$ as output register for the output in $\Bool$;
$\rmreg{2}$ as output register for the output in $\Nat$;
$\rmreg{3}$, $\rmreg{4}$ and $\rmreg{5}$ as auxiliary registers.
The content of $\rmreg{1}$ represents the Boolean output as follows:
$0$ represents $\True$ and all other natural numbers represent $\False$.
For each $i \in [0,5]$, register $\rmreg{i}$ can be
incremented by one, decremented by one, and tested for zero by means of
instructions
$\rmreg{i}.\rmincr$, $\rmreg{i}.\rmdecr$ and $\rmreg{i}.\rmiszero$,
respectively.
We write $\RML$ for the set of all \PGLBbt\ instruction sequences,
taking the set
$\set{\rmreg{i}.\rmincr,\rmreg{i}.\rmdecr,\rmreg{i}.\rmiszero \where
      i \in [0,5]}$
as the set $\BInstr$ of basic instructions.
Clearly, $\RML$ is adequate to represent all register machine programs
using six registers.

We define a computable functional unit $\Univ \in \FU(\Nat)$ whose
method operations can simulate the effects of the register machine
instructions by encoding the register machine states by natural numbers
such that the contents of the registers can reconstructed by prime
factorization.
This functional unit is defined as follows:
\begin{ldispl}
\begin{aeqns}
\Univ & = &
\set{\tup{\expii,\Expii},\tup{\factv,\Factv}}
\\    & \union &
\set{\tup{\rmmn{i}{succ},\rmmo{i}{succ}},
     \tup{\rmmn{i}{pred},\rmmo{i}{pred}},
     \tup{\rmmn{i}{iszero},\rmmo{i}{iszero}} \where
     i \in [0,5]}\,,
\end{aeqns}
\end{ldispl}
where the method operations are defined as follows:
\begin{ldispl}
\begin{aeqns}
\Expii(x) & = & \tup{\True,2^x}\;, \\
\Factv(x) & = &
\tup{\True,\max \set{y \where \Exists{z}{x = 5^y \mmul z}}}
\end{aeqns}
\end{ldispl}
and, for each $i \in [0,5]$:%
\footnote
{As usual, we write $x \divs y$ for $y$ is divisible by $x$.}
\begin{ldispl}
\begin{aeqns}
\rmmo{i}{succ}(x) & = & \tup{\True,\prim_i \mmul x}\;, \\
\rmmo{i}{pred}(x) & = &
\Biggl\{
\begin{array}[c]{@{}l@{\;\;}l@{}}
\tup{\True, x \mdiv \prim_i} & \mif \prim_i \divs x \\
\tup{\False,x}               & \mif \Not (\prim_i \divs x)\;,
\end{array}
\beqnsep
\rmmo{i}{iszero}(x) & = &
\Biggl\{
\begin{array}[c]{@{}l@{\;\;}l@{}}
\tup{\True, x} \phantom{{} \mdiv \prim_i}
                             & \mif \Not (\prim_i \divs x) \\
\tup{\False,x}               & \mif \prim_i \divs x\;,
\end{array}
\end{aeqns}
\end{ldispl}
where $\prim_i$ is the $(i{+}1)$th prime number, i.e.\
$\prim_0 = 2$, $\prim_1 = 3$, $\prim_2 = 5$, \ldots\ .

We define a function $\rmlful$ from $\RML$ to $\Lf{\IF(\Univ)}$, which
gives, for each instruction sequence $P$ in $\RML$, the instruction
sequence in $\Lf{\IF(\Univ)}$ by which the effect produced by $P$
on a register machine with six registers can be simulated on $\Univ$.
This function is defined as follows:
\begin{ldispl}
\rmlful(u_1 \conc \ldots \conc u_k)
\\ \quad {} =
f.\expii \conc \phi(u_1) \conc \ldots \conc \phi(u_k) \conc
\ntst{f.\rmmn{1}{iszero}} \conc \fjmp{3} \conc
f.\factv \conc \haltP \conc f.\factv \conc \haltN\;,
\end{ldispl}
where
\begin{ldispl}
\begin{aceqns}
\phi(a) & = & \psi(a)\;, \\
\phi(\ptst{a}) & = & \ptst{\psi(a)}\;, \\
\phi(\ntst{a}) & = & \ntst{\psi(a)}\;, \\
\phi(u) & = & u
 & \mif u \;\mathrm{is\;a\;jump\;or\;termination\;instruction}\;,
\end{aceqns}
\end{ldispl}
where, for each $i \in [0,5]$:
\begin{ldispl}
\begin{aceqns}
\psi(\rmreg{i}.\rmincr)   & = & f.\rmmn{i}{succ}\;, \\
\psi(\rmreg{i}.\rmdecr)   & = & f.\rmmn{i}{pred}\;, \\
\psi(\rmreg{i}.\rmiszero) & = & f.\rmmn{i}{iszero}\;.
\end{aceqns}
\end{ldispl}

Take an arbitrary computable $M \in \MO(\Nat)$.
Then there exist an instruction sequence in $\RML$ that computes $M$.
Take an arbitrary $P \in \RML$ that computes $M$.
Then $\moextr{\rmlful(P)}{\Univ} = M$.
Hence, $M$ is a derived method operation of $\Univ$.
\qed
\end{proof}
The universal computable functional unit $\Univ$ defined in the proof of
Theorem~\ref{theorem-universal-fu} has $20$ method operations.
However, three method operations suffice.
\begin{theorem}
\label{theorem-universal-fu-three-meths}
There exists a computable $\cH \in \FU(\Nat)$ with only three method
operations that is universal.
\end{theorem}
\begin{proof}
We know from the proof of Theorem~\ref{theorem-universal-fu} that there
exists a computable $\cH \in \FU(\Nat)$ with $20$ method operations, say
$M_0$, \ldots, $M_{19}$.
We will show that there exists a computable $\cH' \in \FU(\Nat)$ with
only three method operations such that $\cH \leq \cH'$.

We define a computable functional unit $\Univiii \in \FU(\Nat)$ with
only three method operations such that $\Univ \leq \Univiii$ as follows:
\begin{ldispl}
\Univiii = \set{\tup{\gi,\Gi},\tup{\gii,\Gii},\tup{\giii,\Giii}}\;,
\end{ldispl}
where the method operations are defined as follows:
\begin{ldispl}
\begin{aeqns}
\Gi(x)   & = & \tup{\True,2^x}\;, \\
\Gii(x)  & = &
\left\{
\begin{array}[c]{@{}l@{\,}l@{}}
\tup{\True, 3 \mmul x}
 & \mif \Not (3^{19} \divs x) \And
        \Forall{y}{(y \divs x \Implies (y = 2 \Or y = 3))} \\
\tup{\True, x \mdiv 3^{19}}
 & \mif 3^{19} \divs x \And \Not (3^{20} \divs x) \And
        \Forall{y}{(y \divs x \Implies (y = 2 \Or y = 3))} \\
\tup{\False,0}
 & \mif 3^{20} \divs x \Or
        \Not \Forall{y}{(y \divs x \Implies (y = 2 \Or y = 3))}\;,
\end{array}
\right.
\beqnsep
\Giii(x) & = & M_{\factiii(x)}(\factii(x))\;,
\end{aeqns}
\end{ldispl}
where
\begin{ldispl}
\begin{aeqns}
\factii(x)  & = & \max \set{y \where \Exists{z}{x = 2^y \mmul z}}\;, \\
\factiii(x) & = & \max \set{y \where \Exists{z}{x = 3^y \mmul z}}\;.
\end{aeqns}
\end{ldispl}

We have that, for each $i \in [0,19]$,
$\moextr
  {f.\gi \conc {f.\gii}^{\,i} \conc
   \ptst{f.\giii} \conc \haltP \conc \haltN}
  {\Univiii} = M_i$.%
\footnote
{For each primitive instruction $u$, the instruction sequence $u^n$ is
 defined by induction on $n$ as follows: $u^0 = \fjmp{1}$, $u^1 = u$ and
 $u^{n+2} = u \conc u^{n+1}$.}
Hence, $M_0$, \ldots, $M_{19}$ are derived method operations of
$\Univiii$.
\qed
\end{proof}
The universal computable functional unit $\Univiii$ defined in the proof
of Theorem~\ref{theorem-universal-fu-three-meths} has three method
operations.
We can show that one method operation does not suffice.
\begin{theorem}
\label{theorem-universal-fu-one-meth}
There does not exist a computable $\cH \in \FU(\Nat)$ with only one
method operation that is universal.
\end{theorem}
\begin{proof}
We will show that there does not exist a computable $\cH \in \FU(\Nat)$
with one method operation such that $\Counter \leq \cH$.
Here, $\Counter$ is the functional unit introduced at the beginning of
this section.

Assume that there exists a computable $\cH \in \FU(\Nat)$ with one
method operation such that $\Counter \leq \cH$.
Let $\cH' \in \FU(\Nat)$ be such that $\cH'$ has one method operation
and $\Counter \leq \cH'$, and let $m$ be the unique method name such
that $\IF(\cH') = \set{m}$.
Take arbitrary $P_1,P_2 \in \Lf{\IF(\cH')}$ such that
$\moextr{P_1}{\cH'} = \Incr$ and $\moextr{P_2}{\cH'} = \Decr$.
Then $\moextr{P_1}{\cH'}(0) = \tup{\True,1}$ and
$\moextr{P_2}{\cH'}(1) = \tup{\True,0}$.
Instruction $f.m$ is processed at least once if $P_1$ is applied to
$\cH'(0)$ or $P_2$ is applied to $\cH'(1)$.
Let $k_0$ be the number of times that instruction $f.m$ is processed on
application of $P_1$ to $\cH'(0)$ and
let $k_1$ be the number of times that instruction $f.m$ is processed on
application of $P_2$ to $\cH'(1)$ (irrespective of replies).
Then, from state $0$, state $0$ is reached again after $f.m$ is
processed $k_0 + k_1$ times.
Thus, by repeated application of $P_1$ to $\cH'(0)$ at most $k_0 + k_1$
different states can be reached.
This contradicts with $\moextr{P_1}{\cH'} = \Incr$.
Hence, there does not exist a computable $\cH \in \FU(\Nat)$ with one
method operation such that $\Counter \leq \cH$.
\qed
\end{proof}
It is an open problem whether two method operations suffice.

\section{Functional Units for Finite State Spaces}
\label{sect-func-unit-fin}

In this short section, we make some remarks about functional units for
finite state spaces.

In the special case where the state space is $\Bool$, the state space
consists of only two states.
Because there are four possible unary functions on $\Bool$, there are
precisely $16$ method operations in $\MO(\Bool)$.
There are in principle $2^{16}$ different functional units in
$\FU(\Bool)$, for it is useless to include the same method operation
more than once under different names in a functional unit.
This means that $2^{16}$ is an upper bound of the number of functional
unit degrees in $\FU(\Bool) \mdiv {\equiv}$.
However, it is straightforward to show that $\FU(\Bool) \mdiv {\equiv}$
has only $12$ different functional unit degrees.

In the more general case of a finite state space consisting of $k$
states, say~$S_k$, there are in principle $2^{2^k \mmul k^k}$ different
functional units in $\FU(S_k)$.
Already with $k = 3$, it becomes unclear whether the number of
functional unit degrees in $\FU(S_k)$ can be determined manually.
Actually, we do not know at the moment whether it can be determined with
computer support either.

\section{Concluding Remarks}
\label{sect-concl}

We have defined the concept of a functional unit for a state space and
have established general results concerning functional units for natural
numbers.
The main result is the existence of a universal computable functional
unit for natural numbers.
The case where the state space is the set of natural numbers is arguably
the simplest significant case.
We have not yet investigated other significant cases.

An interesting case is the one where the state space is the set of all
pairs of sequences over some alphabet: the tape of a Turing machine can
be modelled by a functional unit for this state space.
Each Turing machine can be simulated by means of a functional unit that
corresponds to the tape of the Turing machine and a \PGLBbt\ instruction
sequence that corresponds to the finite control of the Turing machine.
Variations of the Turing machine theme can be dealt with in this way as
well.
Thus, functional units allows for many computability issues to be viewed
as issues about programs rather than machines.

In~\cite{BM09k}, we introduce an extension of program algebra with
Boolean termination instructions, called \PGAbt, and define a thread
extraction operation for it.
\PGLBbt\ instruction sequences can be translated into closed \PGAbt\
terms such that thread extraction for \PGLBbt\ yields the same
behaviours as translation followed by thread extraction for \PGAbt.
In~\cite{BM09k}, we also introduce an extension of basic thread algebra
similar to \TAbt.
In addition to the constants and operators of \TAbt, that extension has
a constant ($\Stop$) for termination without delivery of a Boolean value
and an operator ($\sfuse$) which is concerned with the effects of
service families on threads and therefore produces threads.

\bibliographystyle{splncs03}
\bibliography{TA}

\end{document}